An Empirical Study of the Behaviour of the Sample Kurtosis in Samples from Symmetric Stable Distributions

J. Martin van Zyl

*Department of Actuarial Science and Mathematical Statistics, University of the Free State, Bloemfontein, South Africa*

e-mail: wwjvz@ufs.ac.za

Kurtosis is seen as a measure of the discrepancy between the observed data and a Gaussian distribution and is defined when the 4th moment is finite. In this work an empirical study is conducted to investigate the behaviour of the sample estimate of kurtosis with respect to sample size and the tail index when applied to heavy-tailed data where the 4th moment does not exist. The study will focus on samples from the symmetric stable distributions. It was found that the expected value of excess kurtosis divided by the sample size is finite for any value of the tail index and the sample estimate of kurtosis increases as a linear function of sample size and it is approximately equal to $n(1-\alpha/2)$.

## 1. Introduction

For heavy-tailed distributions, the theoretical kurtosis is defined and finite when the 4th moment is finite or in terms of the tail index, $\alpha$, where $\alpha > 4$. In practice data is observed with an unknown distribution and kurtosis is used to measure how leptokurtic the sample is. In financial data it is often observed that $\alpha < 2$ and the estimated kurtosis is used to get an indication of how leptokurtic the data is. Estimates of kurtosis in asset returns range from 4 to 50 (Engle and Patton, 2001). Heavy-tailed distributions with $\alpha < 4$ is fitted to log-returns, see for example Xu,Wu and Xiao (2011).

In this work an empirical study is conducted to check the behaviour and usefulness of sample kurtosis for symmetric stable distribution with $\alpha \leq 2$, and specifically where $\alpha \geq 1$which is mostly found when applied to real data.

The main result found was that the expected value of the sample kurtosis increases as a linear function of the sample size and it was found that that for symmetric samples from stable distributions, the approximate sample estimate of kurtosis for a sample size $n$ and tail index $\alpha$ is $n(1-\alpha/2)$.

More than one method was suggested to estimate kurtosis but in this work the Pearson kurtosis as discussed by Fiori and Zenga (2009) which is used in finance and risk analysis is used. Kurtosis is defined as

$$\beta_2(x) = E(X-\mu)^4 / (E(X-\mu))^2$$

$$= \mu_4 / \sigma^4, \qquad (1)$$

with $\mu_4$ the fourth central moment and $\sigma^2$ the variance of $X$. $\beta_2(x)$ is location-scale invariant and all data simulated will be for a location parameter $\mu = 0$ and scale parameter $\sigma = 1$. For a regular distribution, $\mu_4 \geq \mu_2^2$, except when the distribution is only concentrated at two points (Kendall, Stuart and Ord (1987, p.107). The excess kurtosis is

$$\gamma_2 = \beta_2 - 3, \qquad (2)$$

which is also equal to $\gamma_2 = \kappa_4 / \kappa_2^2$ when expressed in terms of cumulants. For the normal distribution, the excess kurtosis is zero.

The sample kurtosis is denoted by $b_2$ and the excess kurtosis by $g_2 = b_2 - 3$. Algebraic inequalities which does not depend on distributional properties were derived for the sample kurtosis and it was shown that for a sample of size *n*, $x_1, ..., x_n$, the sample estimate of kurtosis is less than the sample size $n$ (Johnson, Lowe (1979), Cox (2010)) , thus

$$b_2 = \frac{1}{n}\sum_{j=1}^{n}(x_j - \overline{x})^4 / (\frac{1}{n}\sum_{j=1}^{n}(x_j - \overline{x})^2)^2 \qquad (3)$$

$$= n\sum_{j=1}^{n}(x_j - \overline{x})^4 / (\sum_{j=1}^{n}(x_j - \overline{x})^2)^2$$

$$= nc(x_1, ..., x_n)$$

$$\leq n.$$

This inequality shows that the function $c(x_1, ..., x_n) \leq 1$ and the expected value of $E(c(x_1, ..., x_n)) = E(b_2 / n) \leq 1$, would be finite for all distributions, which means that divergence of the sample kurtosis is because of an increase in the sample size. The behaviour of the sample kurtosis will be like that of a ratio, and not by considering the numerator and denominator separately as is done in the theoretical definition. Using simulation studies it was checked if $c(x_1, ..., x_n)$ can be approximated as a function of $\alpha$. It can also be seen and was confirmed using simulation that the variance of the sample kurtosis is of the form $n^2 \operatorname{var}(c(x_1,...,x_n))$.

This work will focus on symmetrical stable distributed data. Properties and applications of it can for example be found in the work of Cizek , Härdle and Weron, eds. (2011).

The characteristic function of the family of stable distributions is denoted by $\phi(t)$ where

$$\log\phi(t) = -\sigma^{\alpha} |t|^{\alpha} \{1 - i\beta sign(t)\tan(\pi\alpha/2)\} + i\mu t,\ \alpha \neq 1,$$

and $$\log\phi(t) = -\sigma|t|\{1 + i\beta sign(t)(2/\pi)\log(|t|)\} + i\mu t,\ \alpha = 1.$$

The parameters are the tail index, $\alpha \in (0,2]$, a scale parameter $\sigma > 0$, coefficient of skewness $\beta \in [-1,1]$ and location parameter $\mu$. The symmetric case with $\beta = 0$ will be considered in this work.

In the following figure $m = 500$ random samples were simulated, $\alpha's$ were randomly chosen on the interval [1,2] and $m = 500$ random sample sizes between $n$=200 and $n$=1500 and the estimated excess kurtosis plotted. The focus of this study is applications in finance and these sample sizes cover 1 to 5 years when working with daily data. To get an idea of the relationship involved, multiple regression was performed and it was found that the relationship is approximately $g_2 \approx n - n\alpha/2$. There is little variation in the regression coefficients when repeating the simulation and this relationship will be investigated further using simulation studies. Assuming that $g_2(n,\alpha) = n(1-\alpha/2)$, and by noting that

$$\frac{\partial g_2(n,\alpha)}{\partial n} = 1 - \alpha/2 \text{ and } \frac{\partial g_2(n,\alpha)}{\partial \alpha} = -n/2,$$

it can be seen that the sample kurtosis is very sensitive with respect changes in $\alpha$ and a slowly increasing function of the sample size.

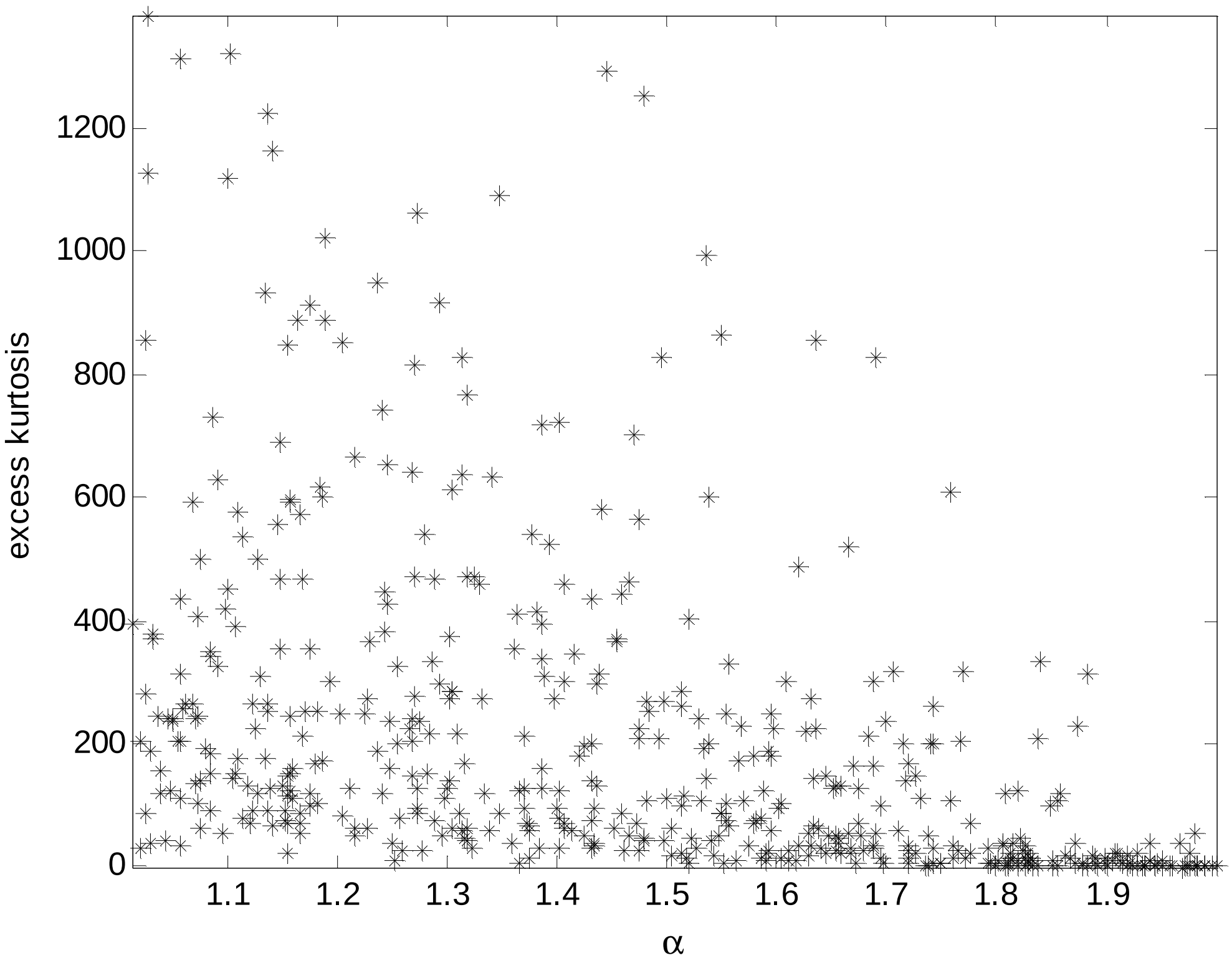

Figure 1. A scatterplot of 500 sample estimates of the excess kurtosis for random $\alpha \in [1,2]$ and the sample size $n$ between 200 and 1500. Samples from a symmetric stable distribution.

The behaviour of sample skewness was checked too using the simulated samples and it was found that the expected value of the sample skewness is zero for symmetric data but the variance is an increasing linear function of the sample size and increases for smaller $\alpha$. This is not the focus of the work, but a skewness estimate in a large sample might not be a significant indication of skewness if the large variance is taken into account.

A measure to order different symmetric distributions according to the term used ‘heavy-tailness’ was derived by van Zwet (1964), Groeneveld and Meeden (1984).  It was proven that if a distribution is more heavy-tailed than another according to this measure, then kurtosis will also be larger for the heavier-tailed distribution.

**2.  Simulation study**

Say a sample of size *n* is available and $n = k_1 + ... + k_r$. The sample kurtosis will be calculated at increasing sample sizes, say $k_1, k_1 + k_2, k_1 + k_2 + k_3, ..., n$, and for different values of $\alpha$.

The following plot shows the rate of increase in the expected value of the estimated excess kurtosis against the number of observations used to calculate it. The slope $\alpha = 1$ for is $b = 0.4917$, for $\alpha = 1.5$ it is $b = 0.2447$ and approximately zero if $\alpha = 2$. The average at each sample size was calculated using $m = 5000$ samples. This relationship can be considered as an approximation. A similar plot where the data is from a student t-distribution with degrees of freedom $\nu = 3, 4$ shows that the relationship for the t-distribution is not linear.

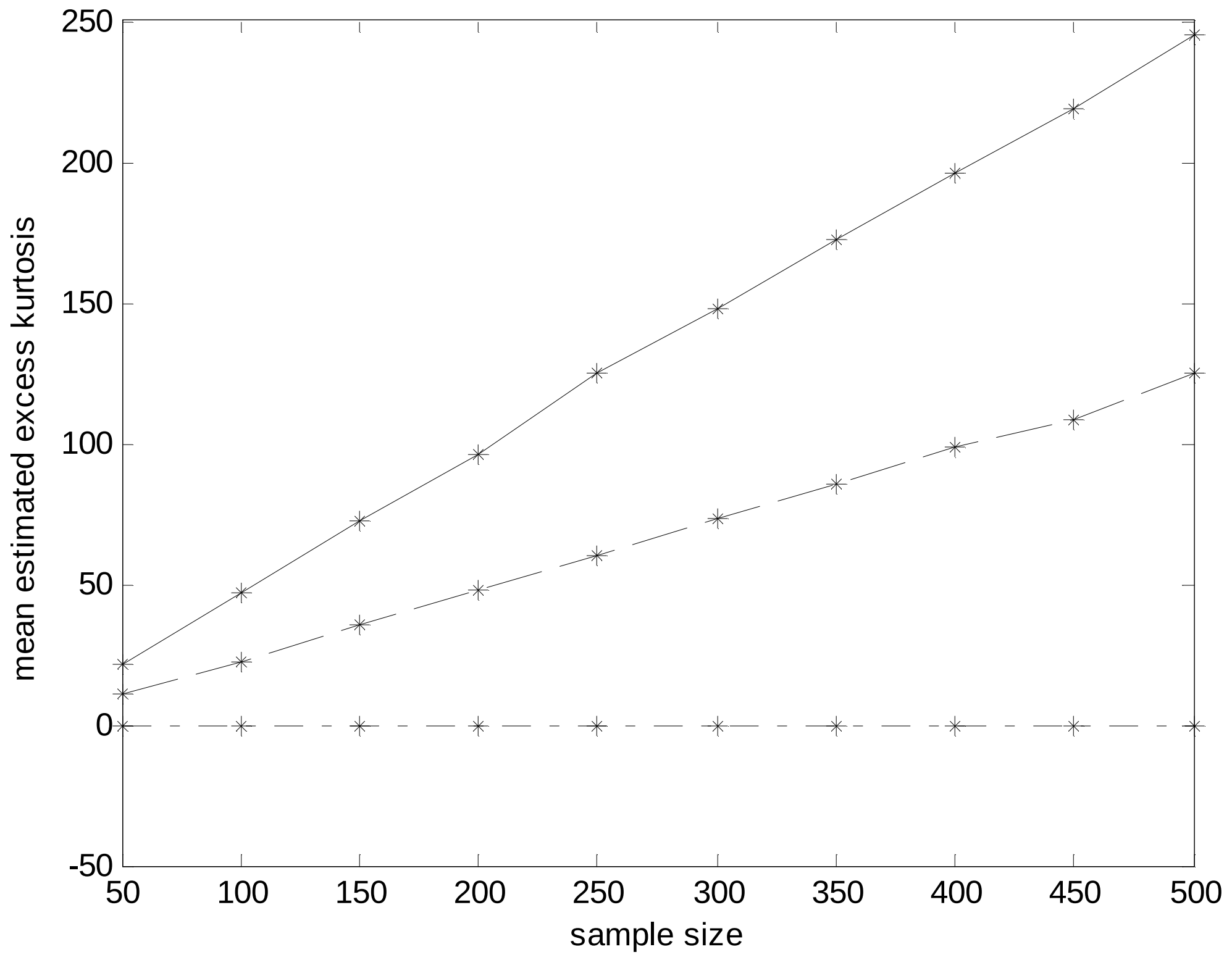

Figure 2. Plot of average of excess sample kurtosis using simulated samples from a symmetric stable distribution. The averages were calculated using 5000 samples, calculated using sample sizes $n = 50, 100, ..., 500$. The solid line is for $\alpha = 1$, dashed line $\alpha = 1.5$ and dash-dot line $\alpha = 2$.

In the following figure the cumulative calculated excess kurtosis is shown for simulated data from a *t*-distribution with $\nu = 3, 4, 5$ degrees of freedom. It can be seen that for small degrees of freedom the relationship between kurtosis and sample size is not linear. The linear trend increase in kurtosis with respect to sample size for sample from a stable distribution can thus be a useful property. It may not be unique but if observed in a practical problem, it means that a possible candidate to fit might be a stable distribution.

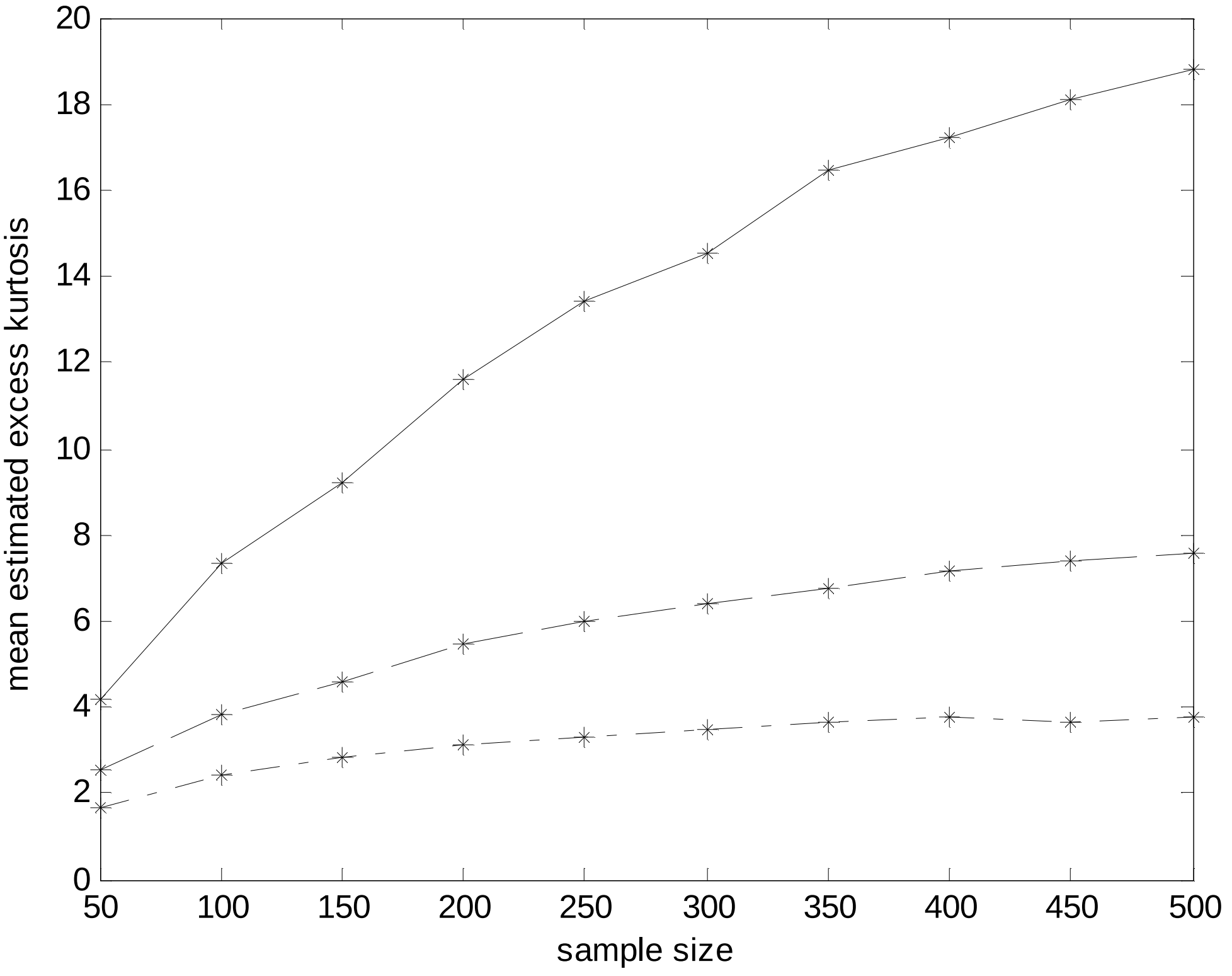

Figure 3. Plot of average of excess sample kurtosis using simulated samples from a t-distribution. The averages were calculated using 5000 samples, calculated using sample sizes $n = 50,100,...,500$. The solid line is for $\nu = 3$, dashed line $\nu = 4$ and dash-dot line $\nu = 5$ degrees of freedom.

For a given value of $\alpha$, denote by $b$ the slope of increase with respect to sample size. If it is assumed that the kurtosis is zero for $\alpha = 2$, regression through the origin can be performed to find the relationship between the slope ($b$) of increase with respect to sample size for a given $\alpha$ and changes in the tail index. As the sample size increases the estimated slope is closer to 2 exactly, leading to the approximate relationship, $b \approx 1 - \alpha / 2$. If this is applied and taking the linear relationship between sample size and sample kurtosis into account, one finds that $g_2 \approx n(1 - \alpha / 2)$.

Figure 3 below is based on the average of 5000 slopes calculated at each $\alpha = 1, 1.1, \ldots, 2$ and fixed sample size $n = 250$.

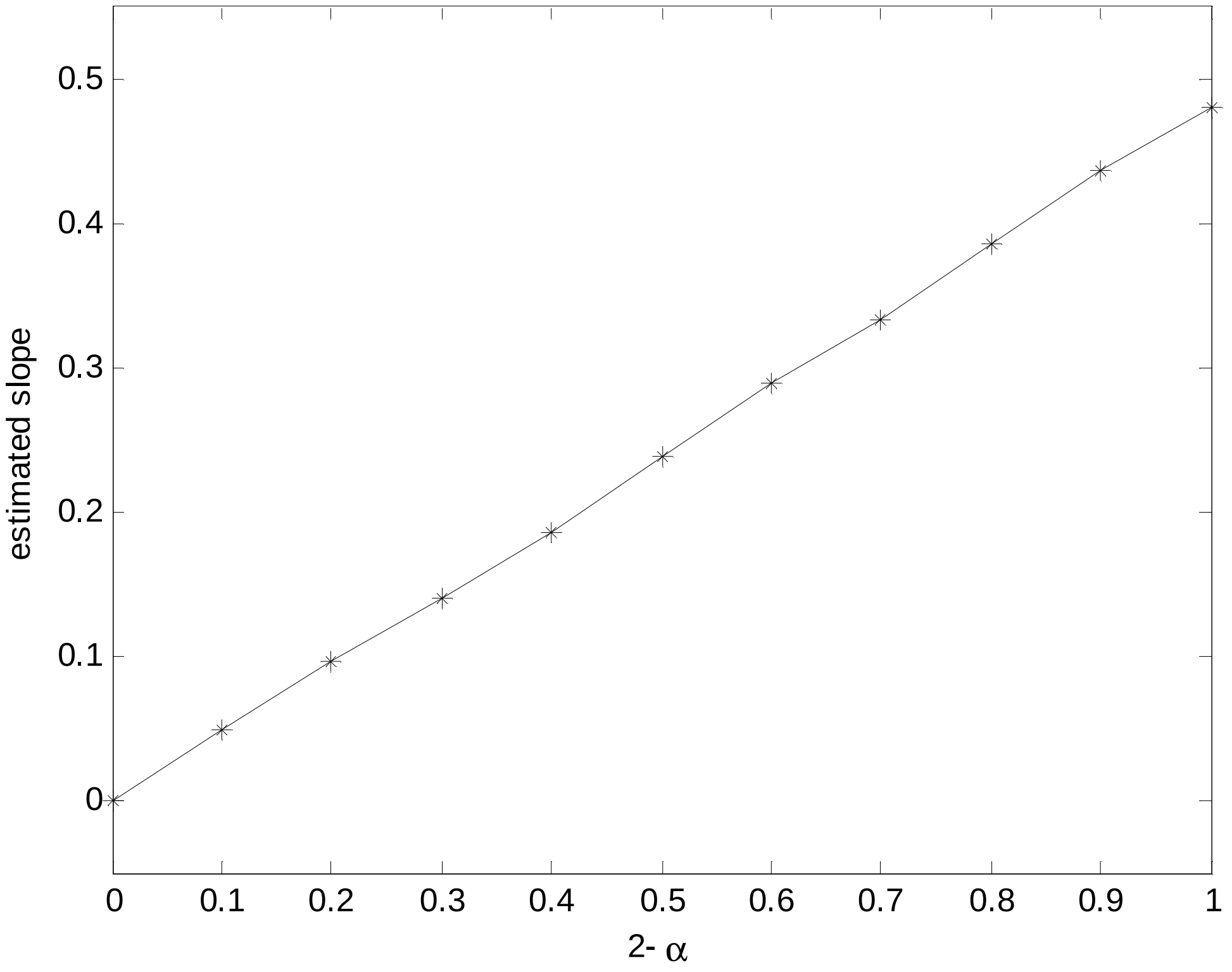

Figure 4. The relationship between the estimated slope of increase of kurtosis with respect to sample size and $2-\alpha$. Each point calculated as the average of 5000 estimated value and $n = 250$.

The expected value of the sample excess kurtosis increases as a linear function of the number of observations used to calculate the sample kurtosis and the approximate expected excess kurtosis in large samples is thus

$$E(g_2) \approx (1-\alpha/2)n. \qquad (5)$$

In the above simulations a fixed sample size was used. To confirm the results it will be checked by using random sample sizes. The consistency of the ordering of kurtosis with tail index will also be considered. As an example 5000 samples with random sample sizes between *n*=200 and *n*=1500 were generated from a stable distribution with index $\alpha = 1.25$, estimated excess kurtosis divided by the sample size and the sample mean was 0.1236 compared to $1-\alpha/2 = 0.1250$. Similarly 5000 samples with random sample sizes and $\alpha = 1.75$ were generated. The sample mean for $\alpha = 1.75$ is 0.3668 compared to $1-\alpha/2 = 0.3750$. This confirms the approximate relationship, $E(g_2) \approx (1-\alpha/2)n$.

Comparisons were made in pairs between the sample kurtosis of the two samples. This resulted in approximately 82% correct larger values of the excess kurtosis when the data is more heavy-tailed. If sample kurtosis was divided by the sample size the percentage increase by about 2%. A few such examples were simulated and when comparing between a normal samples $(\alpha = 2)$ and samples with $\alpha < 2$, the percentage correct ordering of the tail index with respect to kurtosis is very high and often as high as 100%. The conclusion can be made that for symmetric stable distributions, kurtosis is an effective measure to compare the tail-heaviness of samples from two distributions with different parameters.

The variation of the sample estimate is a function of $\alpha$ as seen in figure 5 and the variance for $n$ points used is proportional to $n^2$ for all values of $\alpha$, except when $\alpha = 2$.

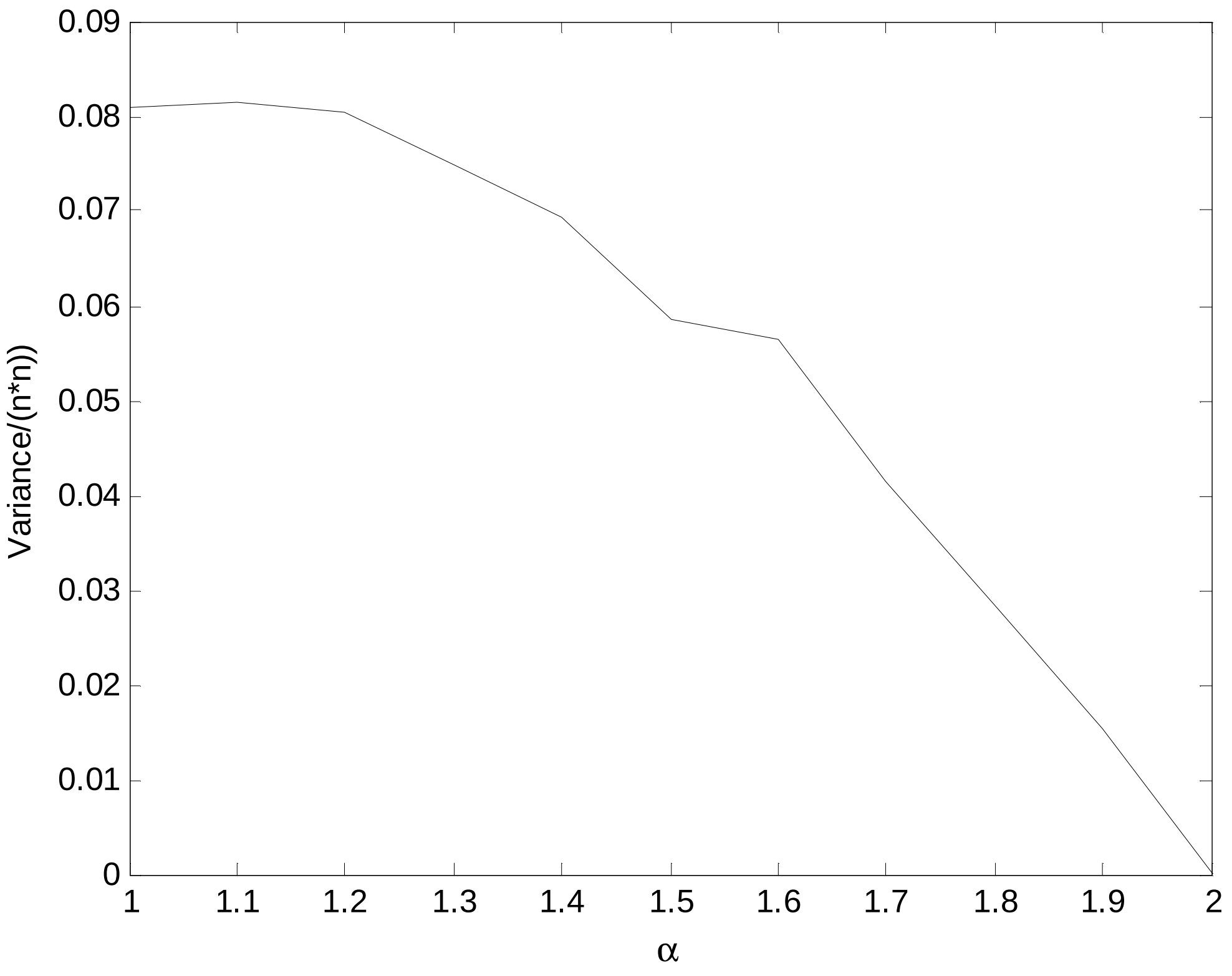

Figure 5. Estimated variance of sample kurtosis divided by $n^2$ of sample of size $n = 500$ as a function of $\alpha$, based on 5000 simulated samples for each $\alpha$.

## 3. An application to log-returns

The change in kurtosis with respect to the number of observations used to calculate the sample kurtosis was investigated when applied to log-returns of the New York stock exchange (NYSE). The daily closing values of 5 years starting in May 2013 to May 2018 were used. Log-returns are approximately symmetrically distributed with sample mean zero and the stable distribution is considered as a possible distribution for log-returns. Log-returns are also approximately independently distributed. The index is shown in figure 6. There is an initial period, a major correction and the period after the correction.

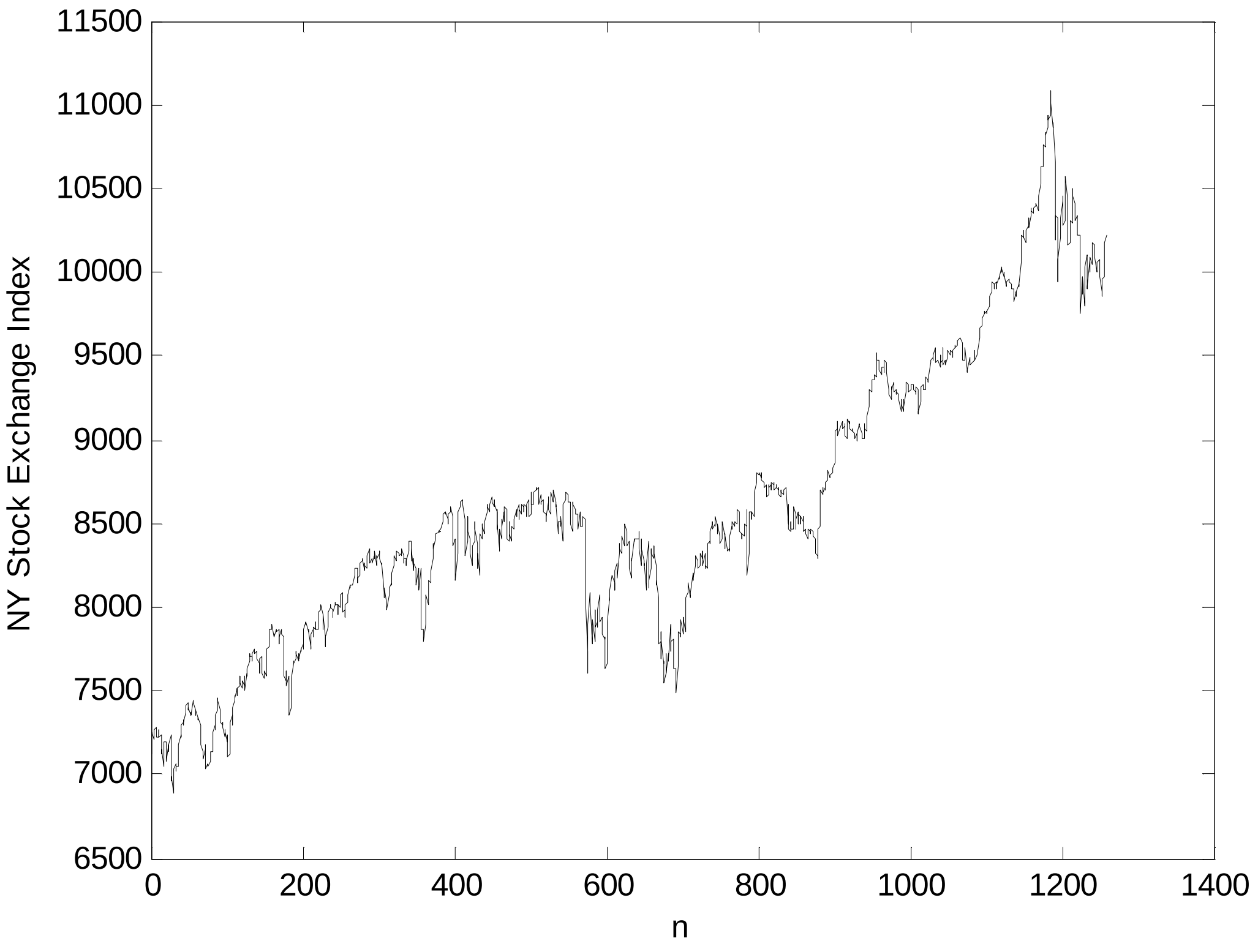

Figure 6. Index of the NY stock exchange, 5 years daily data.

The sample excess kurtosis of the log-returns using increasing sample sizes is plotted in figure 7. It can be seen that the distribution of the log-returns seems to change and then stay the same for a period if one considers a change in slope as an indication of a change in the distribution.

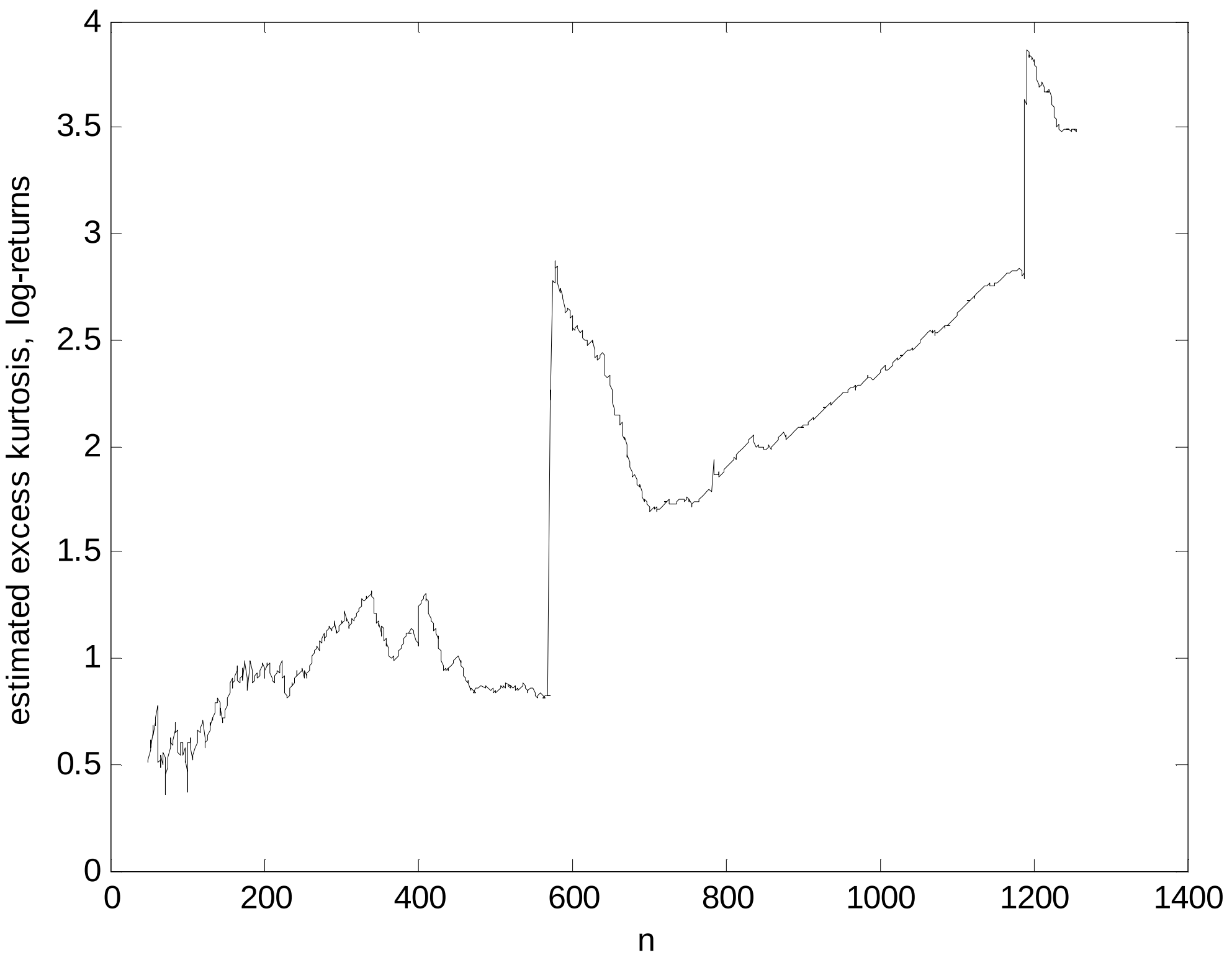

Figure 7. Excess kurtosis of the log-returns calculated as a function of the number of points used to calculate the sample kurtosis.

Kurtosis is very sensitive with respect to changes in the index even though the kurtosis is calculated using log-returns and the difference in behaviour of the sample kurtosis over time and before and after the correction is very clear. The program Stableregkw developed for Matlab by Borak, Misiorek and Weron for the book of Cizek , Härdle and Weron, eds. (2011).based on the Kogon-Williams (Kogon, Williams (1998)) estimation method was applied to two series of observations 100 – 400 and 800 – 1100 to see if there was a change in the tail-index as indicated by the change in sample kurtosis. The estimated tail index for the first period is $\hat{\alpha} = 1.8554$ and $\hat{\alpha} = 1.7165$ for the second period. The estimated parameters when all 1257 log-returns are used is

$\hat{\alpha} = 1.7532,\ \hat{\beta}=0.1184,\ \sigma = 0.0044$.

This is consistent with the change in kurtosis, showing that the second period can be more volatile and heavy-tailed. This is a period during and after an election in the USA.

## 4. Conclusions

There is a relationship between kurtosis and the tail-index for samples from the stable distributions. For a sample of size *n*, the sample kurtosis can be considered as $n$ time the ration of two polynomials which both are of degree 4 and the expected value of the ratio is finite, even if expected value of the numerator or denominator does not exist. This property makes kurtosis useful in heavy-tailed data if the proportionality to $n$ is taken into account. Thus for $\alpha > 0$,

$$\lim_{n\to\infty} \sum_{j=1}^{n} (x_j - \overline{x})^4 / (\sum_{j=1}^{n} (x_j - \overline{x})^2)^2 \approx 1 - \alpha / 2 .$$

This property can be used to compare the 'tail-heaviness' by using kurtosis of two samples from a stable distribution.

The linear relationship between the increase as more points are used to calculate kurtosis can be used as a property to exclude or include a stable distribution as a possible distribution which can be fitted to for example log-returns.

For Garch models the $4^{th}$ moment should be finite when fitted to log-returns. By plotting the estimated kurtosis as a function of an increasing number of observations an increase in sample kurtosis might be an indication that the $4^{th}$ moment is not finite.

Using bootstrap methods to estimate a variance, the relationship $g_2 / n \approx 1-\alpha/2$ can be used in large samples to test hypotheses concerning $\alpha$ and especially to test if $\alpha < 2$.